\UseRawInputEncoding 
\documentclass[pra,twocolumn,shownopacs,amssymb]{revtex4}
\usepackage{graphicx}
\usepackage{amsmath}

\begin{document}

\title{Non-Hermitian BCS-BEC evolution with a complex scattering length}

\author{M. Iskin}
\affiliation{Department of Physics, Ko\c{c} University, Rumelifeneri Yolu, 
34450 Sar\i yer, Istanbul, Turkey}

\date{\today}

\begin{abstract}

Having both elastic and inelastic two-body processes that are characterized 
by a complex $s$-wave scattering length between $\uparrow$ and 
$\downarrow$ fermions in mind, here we apply the non-Hermitian extension 
of the mean-field theory to the BCS-BEC evolution at zero temperature. 
We construct the phase diagram of the system, where we find a reentrant 
superfluid (SF) transition that is intervened by a normal and/or a metastable 
phase as a function of increasing inelasticity. This transition occurs in 
a large parameter regime away from the unitarity, i.e., both on the BCS and 
BEC sides of the resonance, and it is mostly governed by the exceptional points. 
In addition, except for the strongly-inelastic regime, we also show that the SF 
phase can be well-described by the condensation of weakly-interacting bosonic 
pairs in the two-body bound state with a complex binding energy. 

\end{abstract}

\maketitle

\section{Introduction}
\label{sec:intro}

The presence of magnetically-tunable Feshbach resonances in ultracold 
collisions permits the ground state of an SF Fermi gas to evolve from the 
BCS limit of weakly-bound and largely-overlapping Cooper pairs to the 
BEC limit of strongly-bound and smally-overlapping bosonic 
molecules~\cite{ufg, giorgini08, strinati18}. For this purpose, since the main 
objective is to understand the effects of elastic collisions between particles, 
one customarily chooses a purely real scattering length, and tunes both its 
magnitude and sign across the resonance, i.e., first the scattering length 
takes small and negative values in the BCS limit, then it diverges and 
changes sign at the resonance, and then it takes small and positive values 
in the BEC limit. In the case of an $s$-wave resonance, this evolution 
turned out to be a crossover phenomenon without a phase transition 
anywhere in between.

Motivated by the recent works on non-Hermitian Fermionic 
superfluidity~\cite{ghatak18, zhou19, okuma19, yamamoto19}, and 
particularly by Ref.~\cite{yamamoto19} on the Hubbard model with a 
complex-valued interaction strength, here we study the non-Hermitian 
extension of the BCS-BEC evolution with a complex $s$-wave scattering 
length between $\uparrow$ and $\downarrow$ fermions in a continuum 
model, whose real (imaginary) part describes the elastic (inelastic)
processes~\cite{kohler05, chin10}. Our self-consistent mean-field 
theory for the ground state is almost identical to that of Ref.~\cite{yamamoto19}, 
except that we allow not only the SF order parameter but also the chemical 
potential to take complex values. At the expense of this complicacy, 
our number equation becomes purely real, and our theory accurately 
reproduces the two-body physics with a complex binding energy in the 
BEC limit~\cite{munote}. 

Some of our primary findings can be summarized as follows.
By constructing the phase diagram of the continuum model, we first reveal 
a reentrant SF transition that is intervened by a normal and/or a metastable 
phase as a function of increasing inelasticity. In contrast to the lattice 
model where a similar transition is reported only in the weakly-bound 
BCS regime~\cite{yamamoto19}, our model exhibits a reentrant 
transition not only on the BCS side of the resonance but also on the 
strongly-bound BEC side except for the crossover region around unitarity. 
Then, in the weakly-inelastic region, we show that the BEC side can 
be well-described by the condensation of weakly-interacting bosonic pairs 
in the two-body bound state with a complex binding energy. However, the 
physics differs considerably in the strongly-inelastic region, where 
the SF phase is a many-body phenomenon.

The rest of the paper is organized as follows. In Sec.~\ref{sec:mft}, we first 
introduce the non-Hermitian extension of the mean-field Hamiltonian, 
and then obtain the self-consistency equations under the notion of 
biorthogonal quantum mechanics. In Sec.~\ref{sec:numerics}, we present 
the phase diagram of the system, and discuss the self-consistent solutions 
for the SF order parameters and the chemical potentials. The paper ends with 
a brief summary of our findings in Sec.~\ref{sec:conc} and an App.~\ref{sec:app} 
on the use of complex scattering parameters in collision physics.

\section{Mean-Field Theory}
\label{sec:mft}

In this paper, we consider the situation where the contact density-density 
interaction $U$ between $\uparrow$ and $\downarrow$ fermions has an 
imaginary component, i.e., $U = U_R + iU_I$ with $U_R \ge 0$ and 
$U_I \ge 0$~\cite{yamamoto19}. 
The physical motivation for the inclusion of such a term into the effective 
Hamiltonian is due to the inelastic two-body loss processes, and it can 
be derived from the quantum master equation with the proper Limbladian 
operator~\cite{ripoll09, durr09, ashida16, yamamoto19, yoshida19, liu20}. 
While the master equation with the quantum-recycle term describes the 
dissipative dynamics of the system at all times, our effective Hamiltonian 
describes only the short-time dynamics during which the recycle term is 
assumed to be negligible. As proposed in Ref.~\cite{ashida16}, a 
complex-valued interaction can effectively be realized with cold atoms 
through postselection (i.e., projecting out the quantum jumps) by a 
continuous monitoring of the particle number.

\subsection{Mean-Field Hamiltonian}
\label{sec:mfH}

When $U$ is a complex number, the effective mean-field Hamiltonian for 
the stationary Cooper pairs can be written as~\cite{yamamoto19}
\begin{align}
\label{eqn:ham}
H_\mathrm{emf} = \sum_\mathbf{k} 
\left( \begin{array} {cc}
c_{\mathbf{k} \uparrow}^\dagger & c_{-\mathbf{k} \downarrow} 
\end{array} \right)
\left( \begin{array} {cc}
\xi_\mathbf{k} & \Delta \\
\bar{\Delta} & -\xi_\mathbf{k}
\end{array} \right)
\left( \begin{array} {c}
c_{\mathbf{k} \uparrow} \\ 
c_{-\mathbf{k} \downarrow}^\dagger 
\end{array} \right),
\end{align}
where $c_{\mathbf{k} \sigma}^\dagger$ ($c_{\mathbf{k} \sigma}$) creates 
(annihilates) a spin-$\sigma$ fermion with momentum $\mathbf{k}$,
$
\xi_\mathbf{k} = \epsilon_\mathbf{k} - \mu
$
with 
$
\epsilon_\mathbf{k} = \hbar^2 k^2/(2m)
$
the usual free-particle dispersion in continuum and $\mu$ the chemical potential. 
Unlike its Hermitian counterpart, it turns out that $\mu = \mu_R + i \mu_I$ 
must have an imaginary component in order for the number equation to 
take purely real values~\cite{munote}. In addition, the complex parameters 
$\bar{\Delta} \ne \Delta^*$ are the non-Hermitian extension of the SF order 
parameter for pairing. 

In this paper, we are interested in the ground state of the system at zero 
temperature that is based on the notion of biorthogonal quantum mechanics
as follows~\cite{brody14}. First of all, given that 
$
H_\mathrm{emf}^\dagger  \ne H_\mathrm{emf}
$ 
is a non-Hermitian Hamiltonian, its right ground state is not the same as the left 
one. Analogous to the usual BCS theory, one can write
$
|\mathrm{BCS} \rangle = \prod_\mathbf{k} (u_\mathbf{k} + 
v_\mathbf{k} c_{\mathbf{k}\uparrow}^\dagger c_{-\mathbf{k}\downarrow}^\dagger) 
|0\rangle 
$
for the right ground state and
$
\langle \langle \mathrm{BCS} | = \langle 0| 
\prod_\mathbf{k} (u_\mathbf{k} + \bar{v}_\mathbf{k} 
c_{-\mathbf{k}\downarrow} c_{\mathbf{k}\uparrow})
$
for the left one~\cite{yamamoto19}. In accordance with the biorthogonal
formalism, these coefficients must satisfy 
$
u_\mathbf{k}^2 + v_\mathbf{k} \bar{v}_\mathbf{k} = 1
$ 
for every $\mathbf{k}$, so that the inner product
$
\langle \langle \mathrm{BCS} | \mathrm{BCS} \rangle = 1
$
is normalized to unity~\cite{brody14}. This leads to
$
u_\mathbf{k} = \sqrt{(E_\mathbf{k}+\xi_\mathbf{k})/(2E_\mathbf{k})}
$
and
$
v_\mathbf{k} = -\sqrt{\Delta (E_\mathbf{k}-\xi_\mathbf{k}) /(2\bar{\Delta} E_\mathbf{k})}
$
for the right ground state, and to
$
\bar{v}_\mathbf{k} = -\sqrt{\bar{\Delta} (E_\mathbf{k}-\xi_\mathbf{k}) /(2\Delta E_\mathbf{k})}
$
for the left one, where the quasiparticle energy
$
E_\mathbf{k} = \sqrt{\xi_\mathbf{k}^2 + \Delta \bar{\Delta}}
$
is a complex number in general, and may host the so-called exceptional points in 
$\mathbf{k}$ space.

\subsection{Exceptional points}
\label{sec:es}

Unlike the Hermitian Hamiltonians that give rise to a real eigenspectrum in an 
orthonormal eigenspace, the complex eigenspectrum and eigenvectors of non-Hermitian 
Hamiltonians may coalesce into one at the so-called exceptional points in the parameter 
space, i.e., they correspond to the degenerate points in a non-Hermitian 
system~\cite{berry04, heiss04}.
Thus, in sharp contrast to a degeneracy in the real spectrum of Hermitian systems, a 
degeneracy in the complex spectrum of non-Hermitian systems makes the Hamiltonian 
matrix defective, i.e., it does not have a complete basis of eigenvectors, and it is not 
diagonalizable. Since much of the novel properties and applications of non-Hermitian 
systems have been attributed to the presence of these points, next we analyze them 
for our system.

Let us consider a generic two-band Bloch Hamiltonian of a non-Hermitian system 
that is governed by the Hamiltonian matrix
$
H_\mathbf{k} = d_{0\mathbf{k}} \tau_0 + \mathbf{d}_\mathbf{k} \cdot \boldsymbol{\tau},
$ 
where $\tau_0$ is a $2 \times 2$ identity matrix, 
$
\boldsymbol{\tau} = (\tau_x, \tau_y, \tau_z)
$ 
is a vector of Pauli matrices, and
$
\mathbf{d}_\mathbf{k} = \mathbf{d}_{R \mathbf{k}} + i \mathbf{d}_{I \mathbf{k}}
$ 
parametrizes, respectively, the Hermitian and anti-Hermitian terms. The eigenvalues 
of this Hamiltonian matrix can be written as
$
E_{s, \mathbf{k}} = d_{0\mathbf{k}} + s\sqrt{d_{R \mathbf{k}}^2 - d_{I \mathbf{k}}^2 
+ 2i \mathbf{d}_{R \mathbf{k}} \cdot \mathbf{d}_{I \mathbf{k}} },
$
where $s = \pm$, and its exceptional points occur when the conditions
$
d_{R \mathbf{k}}^2 = d_{I \mathbf{k}}^2
$
and
$
\mathbf{d}_{R \mathbf{k}} \cdot \mathbf{d}_{I \mathbf{k}} = 0
$
are simultaneously satisfied~\cite{okugawa19,budich19}. 
Here $d_{R \mathbf{k}}$ and $d_{I \mathbf{k}}$ are the magnitudes of the 
corresponding vectors. For our Hamiltonian, we set
$
d_{0\mathbf{k}} = 0,
$
and choose the gauge
$
\Delta = (\Delta_R + i \Delta_I)e^{i\theta}
$
and
$
\bar{\Delta} = (\Delta_R + i \Delta_I)e^{-i\theta}
$
~\cite{yamamoto19}, leading to
$
\mathbf{d}_{R \mathbf{k}} = ( \Delta_R \cos\theta, -\Delta_R \sin\theta, \epsilon_\mathbf{k} - \mu_R )
$
and
$
\mathbf{d}_{I \mathbf{k}} = ( \Delta_I \cos\theta, -\Delta_I \sin\theta, - \mu_I ).
$
Thus, the exceptional points occur when the conditions
$
\Delta_R^2 + (\epsilon_\mathbf{k}-\mu_R)^2 = \Delta_I^2 + \mu_I^2
$
and
$
\Delta_R \Delta_I = \mu_I(\epsilon_\mathbf{k} - \mu_R)
$
are simultaneously satisfied. Assuming $\mu_I < 0$ (i.e., see our numerical results below), 
these conditions reduce to
$
\mu_I = - \Delta_R
$
and
$
\epsilon_\mathbf{k} = \mu_R - \Delta_I,
$
and they correspond to a surface of exceptional points when $\mu_R > 0$.

\subsection{Self-consistency Equations}
\label{sec:sce}

In terms of the right and left ground states, the SF order parameters can be 
written as the expectation values of the pair annihilation and creation 
operators where
$
\Delta = U \sum_\mathbf{k} \langle \langle
c_{\mathbf{k} \uparrow} c_{-\mathbf{k} \downarrow} 
\rangle 
$
and
$
\bar{\Delta} = U \sum_\mathbf{k} \langle \langle
c_{-\mathbf{k} \downarrow}^\dagger c_{\mathbf{k} \uparrow}^\dagger 
\rangle.
$
They both lead to the order parameter equation
$
1/U = \sum_\mathbf{k} 1/(2E_\mathbf{k})
$
~\cite{yamamoto19}, where 
$
\langle \langle
c_{\mathbf{k} \uparrow} c_{-\mathbf{k} \downarrow} 
\rangle = u_\mathbf{k} v_\mathbf{k}
$
and
$
\langle \langle
c_{-\mathbf{k} \downarrow}^\dagger c_{\mathbf{k} \uparrow}^\dagger 
\rangle = u_\mathbf{k} \bar{v}_\mathbf{k}.
$
Here we follow the usual BCS-BEC crossover approach~\cite{engelbrecht97}, 
and substitute
$
1/U = -mV/(4\pi \hbar^2 a_s) + \sum_\mathbf{k} 1/(2\epsilon_\mathbf{k}),
$
where $V$ is the volume, and the $s$-wave scattering length
$a_s = a_R + i a_I$ between $\uparrow$ and $\downarrow$ fermions 
in vacuum is a complex number with $a_I < 0$ when $U_I > 0$. 
See App.~\ref{sec:app} for their connection to the physical parameters.
Similarly, the number of particles can be obtained from the expectation 
value of the number operator where
$
N = \sum_{\mathbf{k} \sigma} \langle \langle
c_{\mathbf{k} \sigma}^\dagger c_{\mathbf{k} \sigma} 
\rangle.
$
This leads to the number equation
$
N = \sum_\mathbf{k} (1 - \xi_\mathbf{k}/E_\mathbf{k})
$
~\cite{yamamoto19}, where 
$
\langle \langle
c_{\mathbf{k} \sigma}^\dagger c_{\mathbf{k} \sigma} 
\rangle = \bar{v}_\mathbf{k} v_\mathbf{k}.
$
Unless we allow $\mu$ to have complex values, the imaginary component
of $N = N_R + i N_I$ is nonzero in general~\cite{munote}. This may not be 
surprising given that the Hermitian operators do not correspond to physical
observables in the biorthogonal quantum mechanics, causing their expectation 
values to be not necessarily real. 

Noting that the SF order parameters always appear as $\Delta \bar{\Delta}$
in the self-consistency equations, we choose a special gauge above satisfying
$
H_\mathrm{emf}^\dagger = H_\mathrm{emf}^*,
$ 
and set
$
\Delta \bar{\Delta} = \Delta_0^2
$
where
$
\Delta_0 = \Delta_R + i \Delta_I
$
is a complex number~\cite{yamamoto19}. To make further progress, 
we also introduce a simpler notation
$
\xi_\mathbf{k}^2 + \Delta_0^2 = A_\mathbf{k} + iB_\mathbf{k}
$
where
$
A_\mathbf{k} = (\epsilon_\mathbf{k} - \mu_R)^2 + \Delta_R^2 - \mu_I^2 - \Delta_I^2
$
and
$
B_\mathbf{k} = 2\Delta_I \Delta_R - 2\mu_I (\epsilon_\mathbf{k} - \mu_R),
$
and define
$
|E_\mathbf{k}| = (A_\mathbf{k}^2 + B_\mathbf{k}^2)^{1/4}
$
and the principal value
$
\phi_\mathbf{k} = \mathrm{atan2}(B_\mathbf{k}, A_\mathbf{k}) \in (-\pi, +\pi].
$
Note that having a branch cut along the negative $A_\mathbf{k}$ axis~\cite{atan2} 
guarantees a positive value for the real part of $E_\mathbf{k}$.
This allows us to decouple the two complex self-consistency equations 
into four real ones:
\begin{gather}
\label{eqn:gapR}
-\frac{mV a_R}{4\pi \hbar^2 |a_s|^2} =  \sum_\mathbf{k} \left[ 
\frac{\cos(\phi_\mathbf{k}/2)} {2|E_\mathbf{k}|} - \frac{1}{2\epsilon_\mathbf{k}}\right], \\
\label{eqn:gapI}
-\frac{mV a_I}{4\pi \hbar^2 |a_s|^2} =  \sum_\mathbf{k} \frac{\sin(\phi_\mathbf{k}/2)} {2|E_\mathbf{k}|}, \\
\label{eqn:numR}
\frac{k_F^3 V}{3\pi^2} = \sum_\mathbf{k} \left[ 1 - \frac{(\epsilon_\mathbf{k}-\mu_R)\cos(\phi_\mathbf{k}/2) - \mu_I\sin(\phi_\mathbf{k}/2)}
{|E_\mathbf{k}|} \right], \\
\label{eqn:numI}
0 = \sum_\mathbf{k} \frac{(\epsilon_\mathbf{k}-\mu_R)\sin(\phi_\mathbf{k}/2) + \mu_I\cos(\phi_\mathbf{k}/2)}
{|E_\mathbf{k}|}.
\end{gather}
Here, $|a_s| = \sqrt{a_R^2+a_I^2}$ is the magnitude of $a_s$, $N_R$ is set 
to its non-interacting value $k_F^3 V/(3\pi^2)$ with $k_F$ the Fermi wave 
vector, and $N_I$ is set to $0$. 

We note that our formalism recovers the usual BCS-BEC crossover problem 
by construction~\cite{engelbrecht97}, i.e., $\Delta_I \to 0$, $\mu_I \to 0$ 
and $\phi_\mathbf{k} \to 0$ in the limit when $U_I \to 0$ or equivalently 
$1/(k_F a_I) \to -\infty$. The $a_I \to 0^-$ limit has been well-studied in the 
past~\cite{ufg, giorgini08, strinati18}, for which case the mean-field theory 
provides a qualitative understanding of the ground state in the entire range 
of $-\infty < 1/(k_Fa_R) < \infty$. Hoping that the non-Hermitian extension of 
the mean-field theory is also valid, i.e., at least for the weakly-inelastic 
region where $1/(k_Fa_I) \lesssim -5$ if not for the strongly-inelastic region 
where $1/(k_Fa_I) \gtrsim -2$ or the extremely-inelastic limit when 
$1/(k_Fa_I) \to 0^-$, next we resort to a fully numerical approach, and 
analyze the effects of a finite $1/(k_F a_I)$ on the SF properties.

\section{Numerical Results}
\label{sec:numerics}

\begin{figure}[htbp]
\includegraphics[scale=0.75]{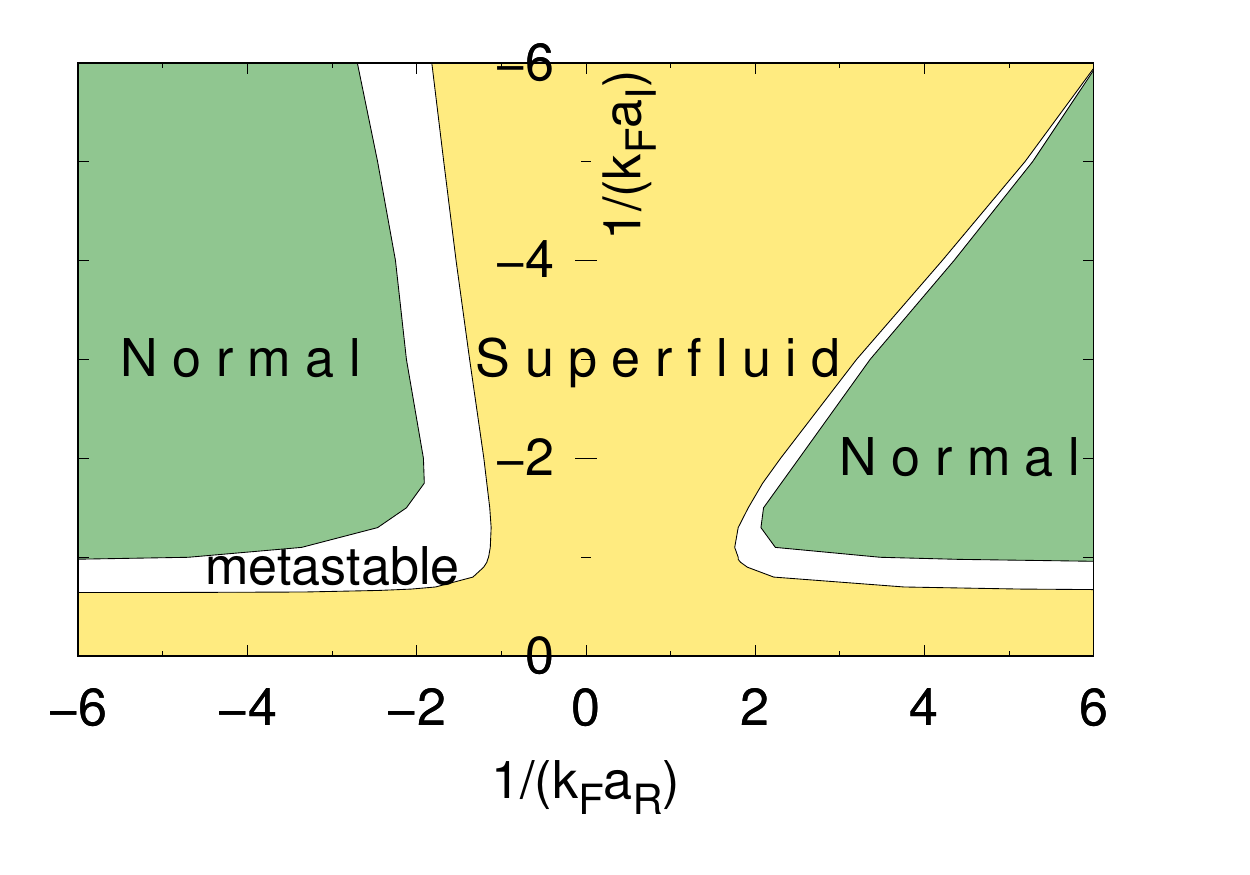}
\caption{
\label{fig:pd} The phase diagram at zero temperature.
In the green-colored normal regions, our numerical calculations do not 
converge reliably to a self-consistent solution with a nontrivial 
$\Delta_R \ne 0$ and/or $\Delta_I \ne 0$. In the white-colored metastable 
regions, the condensation energy of the SF solution does not correspond 
to the global minimum of the real part of the energy, i.e., the nontrivial 
solutions are energetically stable only in the yellow-colored SF region.
}
\end{figure}

After iterating Eqs.~(\ref{eqn:gapR})-(\ref{eqn:numI}) for self-consistent 
solutions of $\Delta_R$, $\Delta_I$, $\mu_R$ and $\mu_I$, we construct 
the phase diagram that is shown in Fig.~\ref{fig:pd}. 
The diagram involves three phases that are characterized by the 
following criteria~\cite{yamamoto19}. In the green regions 
denoted as `Normal', our numerical calculations do not converge to a 
self-consistent solution with a finite $\Delta_R \ne 0$ and/or $\Delta_I \ne 0$. 
While we find convergent solutions with a reliable accuracy in both the 
white regions denoted as `metastable' and the yellow region denoted as 
`Superfluid', these regions are distinguished by the sign of the real part of 
the condensation energy. Here, the condensation energy 
$
E_c = \Delta_0^2/U - \sum_\mathbf{k} (E_\mathbf{k} - \sqrt{\xi_\mathbf{k}^2})
= - \sum_\mathbf{k} (E_\mathbf{k} - \sqrt{\xi_\mathbf{k}^2})^2/(2E_\mathbf{k})
$
corresponds to the energy difference between the SF and normal phases,
and its positive (negative) real part suggests a metastable (stable) SF 
solution, i.e., the SF solution is a local (global) minimum of the real part of
the energy. 

In Fig.~\ref{fig:pd}, we have two disconnected normal regions. The one on 
the BCS side of the resonance, i.e., when $1/(k_F a_R) < 0$, is quite 
similar in structure to the recent work on the lattice model~\cite{yamamoto19}: 
there is a reentrant SF transition that is intervened by a normal and/or a 
metastable phase as a function of increasing $1/(k_Fa_I)$ from $-\infty$ 
towards $0^-$. We find that the conditions $A_\mathbf{k} = 0$ and 
$B_\mathbf{k} = 0$ (or equivalently $\mu_I = - \Delta_R < 0$ with $\mu_R > 0$) 
are simultaneously satisfied on the periphery of the normal region, and therefore, 
the normal region is attributed entirely to the presence of exceptional points in the 
energy spectrum.
When $1/(k_Fa_I) \lesssim -2$, we noticed that $\min A_\mathbf{k}$ 
is negative (positive) in the metastable (SF) region, and that the 
energetic-stability boundary coincides very well with the condition 
$\min A_\mathbf{k} = 0$ or simply $\Delta_R^2 = \Delta_I^2 + \mu_I^2$
in our continuum model. In connection to this, we also observe that the 
momentum distribution, that is given by the summand $[\cdots]$ of
Eq.~(\ref{eqn:numR}), of the metastable phase is not strictly bounded by $0$ 
from below and $2$ from above in a tiny $\mathbf{k}$-space region nearby 
the Fermi momentum. However, curiously enough, the Pauli principle is not 
violated on the energetically-stable side in the SF region.

On the other hand, the normal region on the BEC side of the resonance, 
i.e., when $1/(k_F a_R) > 0$, has no counterpart in the lattice 
model~\cite{yamamoto19}. We believe this difference is quite intuitive 
given the distinct nature and properties of the tightly-bound bosonic pairs 
in these models. In the lattice model, the pairs become strongly repulsive 
when they are on the same site, due to the important role played by the 
Pauli exclusion principle~\cite{iskin08}. In sharp contrast, the pairs 
become weakly repulsive in the continuum model~\cite{engelbrecht97}. 
As the real part of the pair-pair scattering length $a_{p,R} \propto a_R$ 
gets weaker with increasing $1/(k_Fa_R)$, the SF phase eventually 
gives its way to the normal phase once the imaginary part 
$a_{p,I} \propto a_I$ of the pair-pair scattering length dominates 
over $a_{p,R}$. In particular, in the weakly-inelastic
region when $1/(k_Fa_I) \lesssim -5$, we note that the transition from the SF 
phase to the normal one occurs approximately at $1/(k_Fa_R) \approx 1/(k_Fa_I)$ 
without a sizeable metastable region in between. 

This motivates us to study the two-body binding problem with a complex 
$a_s$. Similar to the expression for the usual two-body problem with a 
real $a_s$, the complex binding energy $\varepsilon_b$ of the two-body 
bound state is determined by
$
1/U = \sum_\mathbf{k} 1/(2\epsilon_\mathbf{k} - \varepsilon_b),
$
leading to
$
\varepsilon_b = -\hbar^2/(ma_s^2).
$
Here, we eliminate $1/U$ in favor of $a_s$ via the relation given in
Sec.~\ref{sec:sce}, and perform the integral over real $k$ using the residue 
theorem after going to the complex $k \to z$ plane. Even though the 
final result is identical in mathematical form to the usual problem with 
a real $a_s$, here 
$
\varepsilon_b = \varepsilon_R + i \varepsilon_I
$
is a complex number in general where
$
\varepsilon_R = \hbar^2 (a_I^2 - a_R^2) / (m |a_s|^4)
$
and
$
\varepsilon_I = 2 \hbar^2 a_I a_R / (m |a_s|^4).
$
Therefore, we conclude that a two-body bound state occurs only when $a_R > 0$ 
and $a_R > |a_I|$, leading to $\varepsilon_R < 0$ and $\varepsilon_I < 0$, and its 
lifetime is determined by 
$
\tau_b = - \hbar/(2 \varepsilon_I).
$
See App.~\ref{sec:app} for further discussion.
The absence of a two-body bound state clearly explains why the SF region 
is bounded by $1/(k_Fa_R) < 1/(k_F|a_I|)$ on the BEC side of the resonance
for the weakly-inelastic region when $1/(k_Fa_I) \lesssim -5$. 

\begin{figure}[htbp]
\includegraphics[scale=0.75]{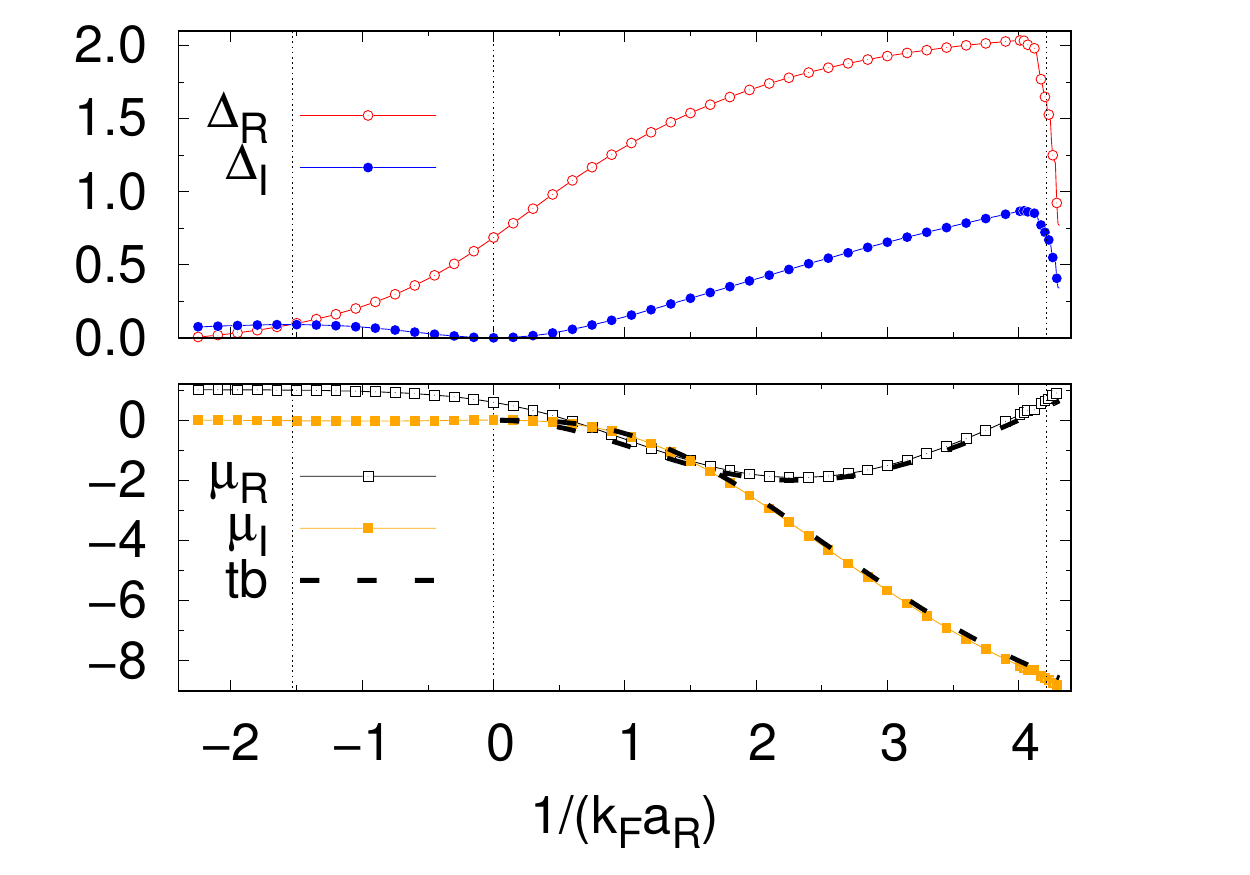}
\caption{
\label{fig:aIm4} 
Self-consistent solutions for $1/(k_Fa_I) = -4$ in units of $\epsilon_F$. 
The vertical lines are guides to the eye for the locations of the phase transitions, 
and the dashed lines are the real and imaginary parts of the two-body result 
$\varepsilon_b/2$.
}
\end{figure}

In order to gain more physical insight into the phase diagram, we set $1/(k_Fa_I)$ 
to $-4$ in Fig.~\ref{fig:aIm4}, and present the resultant self-consistent solutions 
as a function of $1/(k_Fa_R)$. The numerical energy scale is the Fermi energy 
$\epsilon_F = \hbar^2 k_F^2/(2m)$. First of all, independently of the value 
of $1/(k_Fa_I)$, both $\Delta_I$ and $\mu_I$ vanish precisely at the resonance 
when $1/(k_Fa_R) = 0$. For this reason, the evolutions of $\Delta_I$ and 
$\mu_I$ are non-monotonous in the BCS-BEC crossover region. Although 
it is not visible in Fig.~\ref{fig:aIm4}, $\mu_I$ is negative and has the shape 
of an inverted bell curve. In the weakly-inelastic region when 
$1/(k_Fa_I) \lesssim -5$, we find that while $\mu_R \to \epsilon_F$ and 
$\mu_I \to 0^-$ in the BCS limit, they approach to the two-body binding energy 
$\mu_R \to \varepsilon_R/2$ and $\mu_I \to \varepsilon_I/2$ in the BEC limit. 
This is clearly seen in Fig.~\ref{fig:aIm4} where the dashed lines correspond 
to $\mu = \varepsilon_b/2$. Thus, similar to the BEC side of the usual 
BCS-BEC crossover problem~\cite{engelbrecht97}, we conclude that the 
SF phase here can also be well-described by the condensation of 
weakly-interacting bosonic pairs in the two-body bound state. 

\begin{figure}[htbp]
\includegraphics[scale=0.75]{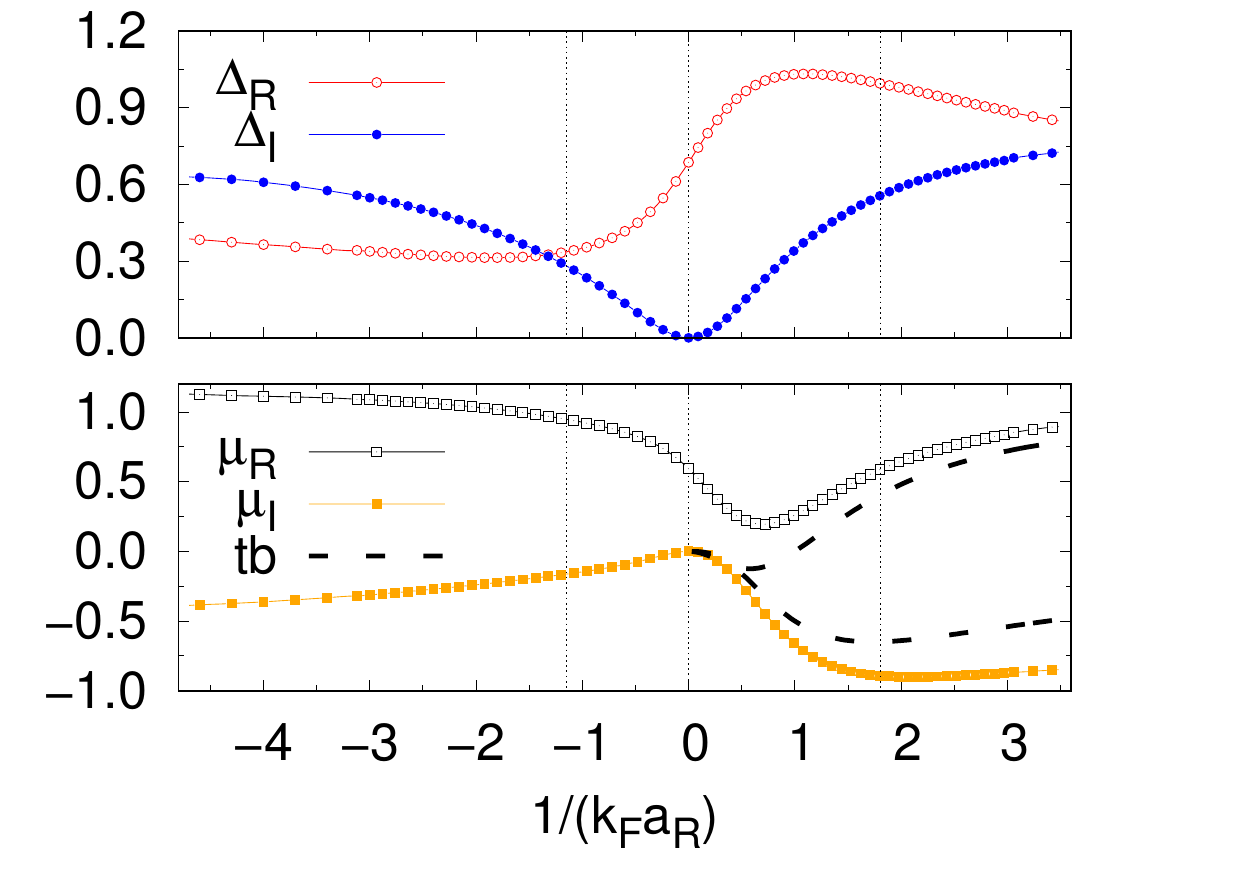}
\caption{
\label{fig:aIm1} 
Self-consistent solutions for $1/(k_Fa_I) = -1$ in units of $\epsilon_F$. 
The vertical lines are guides to the eye for the locations of the phase transitions, 
and the dashed lines are the real and imaginary parts of the two-body result 
$\varepsilon_b/2$.
}
\end{figure}

On the other hand, the physics differs considerably in the strongly-inelastic region
especially when $1/(k_Fa_I) \gtrsim -2$. To illustrate this, we set $1/(k_Fa_I)$ 
to $-1$ in Fig.~\ref{fig:aIm1}, and present the resultant self-consistent solutions 
as a function of $1/(k_Fa_R)$. We find that not only $\mu_R > \epsilon_F$ is 
above the Fermi energy and $\mu_I < 0$ is nonvanishing in the BCS limit, 
they also deviate substantially from the two-body result in the BEC limit. 
We note that, given the absence of a two-body bound state when 
$1/(k_Fa_R) > 1/(k_F|a_I|)$, the SF phase here is a many-body phenomenon 
just like the BCS side. Furthermore, in the extremely-inelastic limit when 
$1/(k_Fa_I) \to 0^-$, we find that $\Delta_R \approx 0.69\epsilon_F$, 
$\Delta_I \to 0$, $\mu_R \approx 0.59\epsilon_F$ and $\mu_I \to 0$ for 
the entire range of $1/(k_Fa_R)$. It suggests that a unitary Fermi SF can be 
achieved by $k_F a_I \to -\infty$ even in the $k_F a_R \to 0^-$ limit.
Whether this amusing result is an indication that the non-Hermitian extension of the 
mean-field theory eventually breaks down in the vicinity of a resonance when 
$k_F a_I \to -\infty$ deserves further investigation. This is because it is analogous 
to the recent observation where strong inelastic collisions were shown to inhibit 
particle losses, and drove a one-dimensional Bose gas into a dissipative but long-lived
strongly-correlated Tonks-Girardeau regime through fermionization of the 
bosons~\cite{syassen08, ripoll09, durr09}. 
While similar results were reported both in continuum and lattice systems, the 
equivalence between strong dissipation and a Pauli exclusion principle was interpreted 
as a manifestation of the continuous quantum Zeno effect only in the lattice 
case~\cite{syassen08, ripoll09}. See also Ref.~\cite{yoshida19} for a similar conclusion
and interpretation in the context of fractional quantum Hall states. 
We also note that the condition $\mu_I = - \Delta_R$ with $\mu_R > 0$ is satisfied 
only in the nearly-flat normal boundary around $1/(k_F a_I) \gtrsim -1$, and therefore, 
only this portion of the boundary can be attributed to the presence of exceptional points 
in the energy spectrum.

\section{Conclusion}
\label{sec:conc}

In summary, here we discussed the non-Hermitian extension of the BCS-BEC
evolution with a complex $s$-wave scattering length $a_s = a_R + ia_I$ 
between $\uparrow$ and $\downarrow$ fermions. Our self-consistent 
mean-field theory for the ground state is almost identical to the recent 
literature~\cite{yamamoto19}, except that we allow not only the SF order 
parameter $\Delta_0$ but also the chemical potential $\mu$ to take complex 
values~\cite{munote}. This turned out to be one of the crucial ingredients of 
the theory in the strongly-bound BEC regime where $2\mu$ approaches 
to the binding energy $\varepsilon_b = -\hbar^2/(ma_s^2)$ of the two-body 
bound state in vacuum. 

Some of our primary findings can be summarized as follows. We 
constructed the phase diagram of the system, where we found a reentrant 
SF transition that is intervened by a normal and/or a metastable phase as a 
function of increasing $1/(k_Fa_I)$ from $-\infty$ towards $0^-$. This transition
occurs in a large parameter window of $1/(k_Fa_R)$ away from the unitarity, 
i.e., both on the BCS and BEC sides of the resonance, and it is entirely 
(partially) governed by the exceptional points when $a_R < 0$ ($a_R > 0$).
Furthermore, in the weakly-inelastic region when $1/(k_Fa_I) \lesssim -5$, 
we showed that the BEC side of the resonance can be well-described by the 
condensation of weakly-interacting bosonic pairs in the two-body bound state 
with a complex $\varepsilon_b$. However, the physics differs considerably in the 
strongly-inelastic region especially when $1/(k_Fa_I) \gtrsim -2$, where 
the SF phase is a many-body phenomenon reminiscent of the BCS side.
As an outlook, the validity of the mean-field theory deserves particular 
investigation in the extremely-inelastic limit when $1/(k_Fa_I) \to 0^-$. 
In addition, it is important to go beyond the non-Hermitian Hamiltonian,
and consider the full Lindblad dynamics of the dissipative SF through, e.g., 
a time-dependent mean-field formalism~\cite{yamamoto20}.

\begin{acknowledgments}
The author thanks A. L. Suba\c{s}{\i} for discussions, and acknowledges 
funding from T{\"U}B{\.I}TAK Grant No. 11001-118F359.
\end{acknowledgments}

\appendix

\section{Resonance scattering with a complex scattering length}
\label{sec:app}

Complex scattering lengths have been widely used in scattering theory, and 
they can be achieved and even measured in multi-channel scattering or optical 
Feshbach resonances. However, it turns out that the resonance becomes much 
less prominent when the imaginary part becomes strong, limiting the tunability 
of the real part of the scattering length. See Sec.II.A.3 in Ref.~\cite{chin10} for 
a detailed discussion of the theory and the analysis of the experimental results.

In particular, our theoretical parameters $a_R$ and $a_I$ can be directly related 
to the physical ones that are discussed in Ref.~\cite{chin10} as follows.
In terms of the collision parameters, the complex $s$-wave scattering length 
$a_s$ is parametrized as
\begin{align}
a_s = a - i b = a_{bg} + a_{bg} \frac{\Gamma_0}{-E_0 + i\gamma/2}
\end{align}
with the imaginary component $b > 0$ characterizing the inelastic scattering 
of atoms. Here $a_{bg}$ is the background scattering length representing
the off-resonant value, 
$\Gamma_0$ is the resonance strength,
$E_0$ is the threshold resonance position, and
$\gamma/\hbar$ is the decay rate for the decay of the bound state into all available
loss channels in the $k \to 0$ limit. The sign of $\Gamma_0$ is the same as the sign
of $a_{bg}$ and it can be both positive or negative.
This leads to~\cite{chin10} 
\begin{align}
a = a_{bg} - a_{bg}  \frac{\Gamma_0 E_0}{E_0^2 + \gamma^2/4}, \\
b = a_{bg}  \frac{\Gamma_0 \gamma} {2E_0^2 + \gamma^2/2},
\end{align}
showing that $a$ attains its maximum variation of $a_{bg} \pm a_{bg} \Gamma_0/\gamma$ 
at $E_0 = \pm \gamma/2$, where $b = a_{bg} \Gamma_0/\gamma$. 
While a quasi-bound state occurs when $a > a_{bg}$ or $E_0 < 0$, it may not be 
possible to tune $a$ at will when $\gamma$ is not weak~\cite{chin10}. 
This is because as the resonance is induced by the bound state getting close to 
the scattering threshold, it becomes much less prominent given that the bound 
state is no-longer well defined. In addition, $b$ attains its peak value when 
$a - a_{bg}$ changes sign~\cite{chin10}.

On the other hand, the decay rate $\gamma$ can be written as 
\begin{align}
\gamma = \frac{2a_{bg} \Gamma_0 b} {(a - a_{bg})^2 + b^2},
\end{align}
showing that $\gamma$ attains (for a given $a - a_{bg}$) its maximum value of 
$a_{bg} \Gamma_0/b$ at $b = a - a_{bg}$. In particular, we note that while $b \ne 0$ 
is the origin of the dissipation of the bound state, a large $b$ does not lead to a 
strong dissipation. More importantly, by reexpressing it as 
$
\gamma = -(2 a_{bg}^2 \Gamma_0^2/E_0) b (a - a_{bg}) / [(a - a_{bg})^2 + b^2]^2,
$
and then matching this expression with the inverse lifetime
$
\hbar/\tau_b = -2\varepsilon_I = -4 \hbar^2 a_I a_R / [m (a_R^2 + a_I^2)^2]
$
of our two-body bound state, we find that $a_R$ plays the role of $a - a_{bg}$ and 
$a_I$ plays the role of $-b$ in the region when $a - a_{bg} > 0$. This is indeed 
consistent with our finding that $a_R > 0$ is one of the requirements for the 
creation of a two-body bound state. 

Even though we introduced the theoretical parameters $a_R$ and $a_I$ to 
characterize the imaginary interaction strength between particles in our effective 
non-Hermitian Hamiltonian, and varied them freely in constructing the resultant 
phase diagram of the system, it is clear that these parameters are not independent 
from each other in cold-atom collisions. 
Given that Fig.~\ref{fig:pd} covers the entire phase space, we hope the 
correspondence discussed above may be used to gain physical intuition about 
physical systems once the relevant parameters of a certain setup are specified.


\begin{thebibliography}{99}
\bibitem{munote} Even in the case of a half-filled lattice model that is considered 
in~\cite{yamamoto19}, our numerical calculations show that setting the real
chemical potential $\mu$ to $0$ gives a nonzero, but negligible, imaginary 
contribution to the number equation. Having a real $\mu$ turns out to be 
problematic in the case of our continuum model, and we circumvent around 
it by allowing a complex $\mu$. In return, our theory captures the two-body 
physics with a complex binding energy when the fermion pairs form tightly-bound
bosonic molecules.
\end{thebibliography}
\end{document}